\documentclass[12pt]{iopart}
\usepackage{iopams}  
\usepackage{graphicx}  
\usepackage{color}  

\begin{document}

\title{Heat dissipation and its relation to thermopower in single-molecule junctions} 

\author{L A Zotti$^1$, M B\"urkle$^2$, F Pauly$^3$, W Lee$^4$, K. Kim$^4$, W Jeong$^4$,
Y Asai$^2$, P Reddy$^{4,5}$ and J C Cuevas$^6$}

\address{$^1$ Departamento de F\'{\i}sica de la Materia Condensada,
Universidad Aut\'onoma de Madrid, E-28049 Madrid, Spain}

\address{$^2$ Nanosystem Research Institute (NRI) ``RICS", National Institute of Advanced 
Industrial Science and Technology (AIST), Umezono 1-1-1, Tsubuka Central 2, Tsukuba, Ibaraki
305-8568, Japan}

\address{$^3$ Department of Physics, University of Konstanz, D-78457 Konstanz, Germany}

\address{$^4$ Department of Mechanical Engineering, University of Michigan, Ann Arbor, 
Michigan, 48109, USA}

\address{$^5$ Department of Materials Science and Engineering, University of Michigan, 
Ann Arbor, Michigan, 48109, USA}

\address{$^6$ Departamento de F\'{\i}sica Te\'orica de la Materia Condensada and
Condensed Matter Physics Center (IFIMAC), Universidad Aut\'onoma de Madrid, E-28049 Madrid, Spain}
\ead{juancarlos.cuevas@uam.es}

\begin{abstract}
Motivated by recent experiments, we present here a detailed theoretical analysis of the Joule heating 
in current-carrying single-molecule junctions. By combining the Landauer approach for quantum transport
with \emph{ab initio} calculations, we show how the heating in the electrodes of a molecular junction is
determined by its electronic structure. In particular, we show that in general the heat is not
equally dissipated in both electrodes of the junction and it depends on the bias polarity 
(or equivalently on the current direction). These heating asymmetries are intimately related to
the thermopower of the junction as both these quantities are governed by very similar principles. We
illustrate these ideas by analyzing single-molecule junctions based on benzene derivatives
with different anchoring groups. The close relation between heat dissipation and thermopower
provides general strategies for exploring fundamental phenomena such as the Peltier effect or 
the impact of quantum interference effects on the Joule heating of molecular transport junctions.
\end{abstract}

\maketitle

\section{Introduction}

The advent of experimental techniques like the scanning tunneling microscope and break junctions
have enabled the creation of single-molecule junctions and the study of their transport properties.
This has triggered the hope to use single molecules as buiding blocks in novel nanoscale electronic
devices, devices that could take advantage of the unlimited physical properties of molecules. In turn, 
this has also given rise to the research field of \emph{Molecular Electronics} \cite{Cuevas2010}. 
Although there are still many basic experimental challenges to be resolved before molecular electronics 
can become a viable technology, it is clear by now that molecular junctions constitute an excellent 
playground to test basic concepts of quantum transport, which could then be applied to novel charge 
and energy nanoscale devices. Thus for instance, the exhaustive study of electronic transport
in single-molecule junctions has led to remarkable progress in the understanding of the basic 
mechanisms that govern the electrical conduction at the molecular scale (for a recent review,
see Ref.~\cite{naturenano-review}). 

In spite of this progress, basic aspects such as thermoelectrical properties and heat dissipation and
transport in molecular junctions have remained largely unexplored due to experimental challenges 
\cite{Dubi2011}. A first step towards this goal was achieved a few years ago with the first studies 
of thermoelectricity in molecular junctions \cite{Reddy2007}. In particular, it has been shown that 
the thermopower provides useful information not contained in the standard current-voltage measurements 
\cite{Paulsson2003}. Intense experimental \cite{Reddy2007,Baheti2008,Malen2009,Tan2010,Yee2011,Tan2011,
Widawsky2011,Evangeli2013,Widawsky2013} and theoretical \cite{Paulsson2003,Pauly2008b,Ke2009,Finch2009,
Liu2009,Bergfield2009,Bergfield2010,Nozaki2010,Quek2011,Sergueev2011,Stadler2011,Saha2011,Bilan2012,
Buerkle2012,Markussen2013} study of the thermopower in molecular junctions have deepened our understanding 
of the charge and energy transport in these systems.

The next step towards quantitative exploration of heat dissipation and conduction at the molecular scale 
has been taken very recently \cite{Lee2013}. In that work we demonstrated that it is possible
to measure the heat dissipated in the electrodes of an atomic-scale junction as a result of the
passage of an electrical current (Joule heating). Further, we also highlighted the novel heat 
dissipation effects that arise in nanoscale systems. To elaborate, while in macroscopic wires the heat 
dissipation is volumetric, it is not obvious \emph{a priori} how the heat is dissipated in devices of 
nanometric size, including molecular junctions. In most of these junctions the inelastic mean free path 
is larger than the characteristic device dimensions and the electrical resistance is dominated by elastic 
processes. Thus, almost all the heat dissipation must take place inside the electrodes at a distance from 
the junction on the order of the inelastic scattering length. Using novel scanning tunneling probes with 
integrated nanothermocouples, we have been able to show that the power dissipated in the electrodes of a 
molecular junction is controlled by its transmission characteristics. In general, heat is not equally 
dissipated in both electrodes and it also depends on the bias polarity, \emph{i.e.}\ on the direction
of the current.

The goal of this work is to explore in more detail the basic principles that govern the
heat dissipation in molecular junctions. For this purpose, we present here a detailed theoretical
analysis of the heat dissipation in these systems based on the combination of the Landauer 
approach with \emph{ab initio} calculations. We pay special attention to the relation between 
the heating asymmetries in single-molecule junctions and their thermopower. As test systems, 
we study junctions based on benzene derivatives that bind to gold electrodes 
via different anchoring groups. For these junctions, we compare the results obtained 
with a transport method based on density functional theory (DFT) with those obtained with 
the so-called DFT+$\Sigma$ approach \cite{Quek2007}, which has been recently introduced to 
correct some of the known deficiencies of DFT-based approaches applied to transport problems. 
The study presented here provides a clear guide on how to control the heat dissipation in 
nanoscale systems and it outlines strategies to explore fundamental issues like the Peltier 
effect in single-molecule junctions.

The rest of the paper is organized as follows. In section \ref{sec-LT} we present a detailed
discussion of the Landauer theory of heat dissipation in nanoscale conductors. In particular,
we discuss the basic formulas and the physical picture underlying heat dissipation in these
systems. We also present some general considerations and establish the connection between heat 
dissipation and thermopower. Moreover, we illustrate our ideas with the help of a toy model 
for molecular junctions. Section \ref{sec-abinitio} is devoted to a description of the two \emph{ab 
initio} methods that we have employed, in combination with the Landauer formalism, to describe 
the heat dissipation and the thermopower in single-molecule junctions. In section \ref{sec-MJ} we 
present a detailed analysis of the heat dissipation and thermopower of single-molecule junctions
based on benzene derivatives with different anchoring groups. Finally, we summarize in section 
\ref{sec-conclusions} the main conclusions of our work.

\section{Landauer theory of heat dissipation} \label{sec-LT}

\subsection{Heat dissipation: Basic formulas and physical picture}

In the late 1950's Rolf Landauer \cite{Landauer1957} put forward a very intuitive quantum mechanical 
approach to describe the electronic transport in nanocircuits that has become one of the central 
pillars of theoretical nanoelectronics \cite{Datta1995}. The key idea of this approach is that if 
the electron transport in a nanojunction is dominated by elastic processes, it can be modeled as 
a scattering problem. In this problem the junction electrodes are assumed to be ideal reservoirs 
where electrons thermalize and acquire a well-defined temperature, while the central region
plays the role of a scattering center. As a result, all the transport properties of these 
systems are completely determined by the transmission function $\tau(E,V)$, which describes
the total probability for electrons to cross the junction at a given energy $E$ when a 
voltage bias $V$ is applied across the system. Landauer's original approach was extended
by several authors to describe both thermal transport and thermoelectricity in nanojunctions
\cite{Sivan1986,Streda1989,Butcher1990,vanHouten1992}, and in this section we explain 
how this approach can be used to describe heat dissipation in atomic-scale circuits.

Let us consider a two-terminal device with source $S$ and drain $D$ electrodes linked by a 
nanoscale junction \cite{note1}. If the electron transport is elastic, \emph{i.e.}\ if electrons 
flow through the device without exchanging energy in the junction region, then obviously their 
excess energy --acquired by the application of a bias voltage-- must be released in the 
electrodes. From simple thermodynamical arguments, the amount of heat released per unit of time 
in an electrode with electrochemical potential $\mu$ is given by \cite{Giazotto2006}
\begin{equation}
Q = \frac{\mu}{e} I - I_E ,
\end{equation}
where $I$ and $I_E$ are the charge and energy current, respectively, and $e>0$ is the electron
charge. Within the Landauer theory, the electronic contribution to charge and energy 
currents in a two terminal device are expressed in terms of the transmission function as
\begin{eqnarray}
\label{eq-Ie}
I(V) & = & \frac{2e}{h} \int^{\infty}_{-\infty} \tau(E,V) [ f_S(E,\mu_S) - f_D(E,\mu_D) ] dE \\
I_E(V) & = & \frac{2}{h} \int^{\infty}_{-\infty} E \tau(E,V) [ f_S(E,\mu_S) - f_D(E,\mu_D) ] dE ,
\label{eq-IE}
\end{eqnarray}
where $f_{S,D}$ are the Fermi functions of the source or the drain and $\mu_{S,D}$ 
are the corresponding electrochemical potentials such that $\mu_S -\mu_D = eV$. The prefactor 2 is due 
to the spin degeneracy that will be assumed throughout our discussion. Using equations 
(\ref{eq-Ie}) and (\ref{eq-IE}) we arrive at the following expressions for the power (heat per unit 
of time) dissipated in the source and the drain electrodes
\begin{eqnarray}
\label{eq-Qs}
Q_S(V) & = & \frac{2}{h} \int^{\infty}_{-\infty} (\mu_S -E) \tau(E,V) [ f_S(E,\mu_S) - f_D(E,\mu_D) ] dE \\
Q_D(V) & = & \frac{2}{h} \int^{\infty}_{-\infty} (E -\mu_D) \tau(E,V) [ f_S(E,\mu_S) - f_D(E,\mu_D) ] dE .
\label{eq-Qd}
\end{eqnarray}
Obviously, the sum of the powers dissipated in both electrodes must be equal to the total power
injected in the junction $Q_{\rm Total} = IV$, as can be seen from equations (\ref{eq-Qs}) and (\ref{eq-Qd})
\begin{equation}
\label{eq-Qtotal}
\fl Q_S(V) + Q_D(V) = \frac{2eV}{h} \int^{\infty}_{-\infty} \tau(E,V) [ f_S(E,\mu_S) - f_D(E,\mu_D) ] 
dE = IV = Q_{\rm Total}(V) .
\end{equation}

\begin{figure}[t]
\begin{center} \includegraphics[width=0.62\columnwidth,clip]{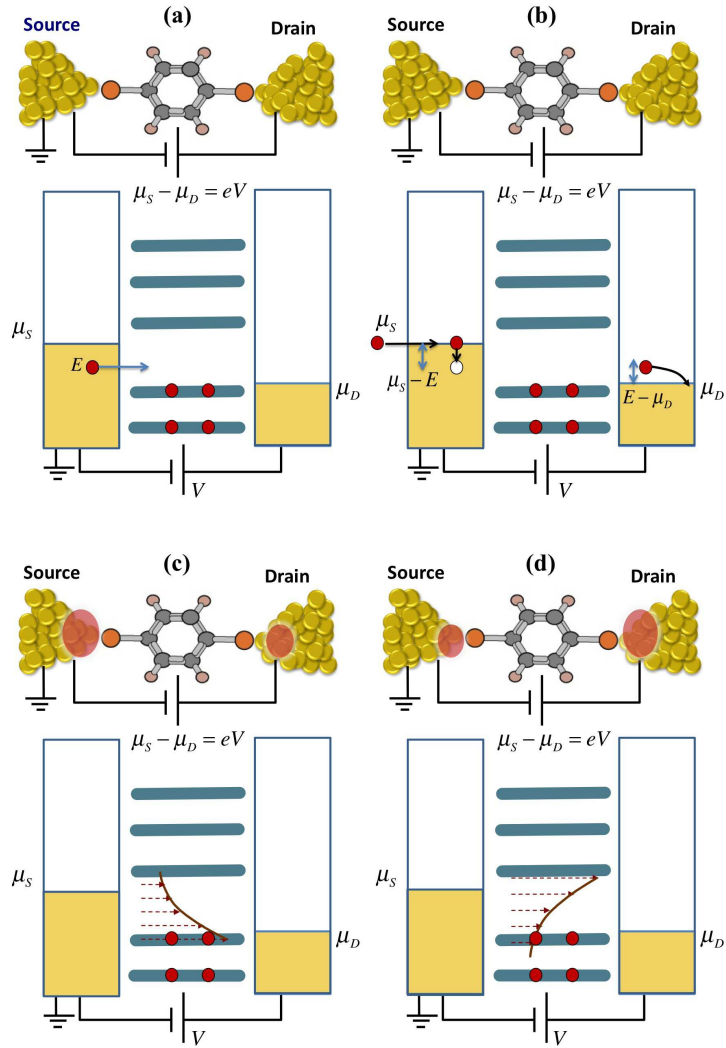} \end{center}
\caption{(a) Schematic representation of the energy diagram of a molecular junction. The source and 
drain are described by two Fermi seas with electrons occupying states up to their respective electrochemical 
potentials. The molecular region is described by a set of discrete state occupied up the HOMO. When 
a bias voltage $V$ is applied across the junction, the electrochemical potential of the source and 
drain shift such that $\mu_S - \mu_D = eV$. (b) When an electron of energy $E$ tunnels from the 
source to the drain it leaves a hole behind. The electron releases its excess energy $E-\mu_D$ in
the drain, while the hole is filled up dissipating an energy equal to $\mu_S-E$ in the source. (c)
If the transmission is energy-dependent and, in particular, if it is higher in the lower part of the 
transport window, then more power is dissipated in the source for positive bias. The transmission function
is represented here by a solid line, while the arrows indicate the direction of the electron flow. 
The red shaded areas help to visualize the different amount of heating in the electrodes. (d) Alternatively, 
if the transmission is higher in the upper part of the transport window, more heat is dissipated in 
the drain for positive bias.} 
\label{fig_HD}
\end{figure}

Equations (\ref{eq-Qs}) and (\ref{eq-Qd}) are the central results of the Landauer theory of heat dissipation
and will be used throughout this work. Given their revelance, it is worth explaining the underlying
physical picture. For this purpose, we shall follow here Ref.~\cite{Datta2005}. As shown 
in figure~\ref{fig_HD}(a) for the case of a molecular contact, when a bias voltage is applied to a junction, 
the electrochemical potentials of the source and the drain are shifted. This opens up an energy window
for electrons to cross the junction and it results in a net electron current in the system. 
Let us consider an electron with an energy $E$ which is transmitted elastically through the junction region 
from the source to the drain. After reaching the drain, see figure~\ref{fig_HD}(b), the electron undergoes
inelastic processes (via electron-phonon interactions) and it decays to the electrochemical 
potential of the drain. In these processes, the electron releases an energy $E-\mu_D$ in the drain. 
On the source side, the original electron leaves behind a hole that is filled up by the same type of 
inelastic processes. In this way, an energy $\mu_S-E$ is dissipated in the source. Notice that the 
overall energy dissipated in the electrodes is equal to $\mu_S - \mu_D = eV$, which is exactly the 
work done by the external battery on each charge. 

Equations (\ref{eq-Qs}) and (\ref{eq-Qd}) describe in quantitative terms the argument just expained. The
energy integrals take into account the contribution of all the electrons in the transport window,
the transmission function accounts for the finite probability of an electron to tunnel through the
junction, and the Fermi functions account for the occupation probabilities of the initial and
final states in the tunneling events. In particular, the appearance of the difference of the Fermi
functions is a result of the net balance between tunneling processes transferring electrons
from the source to the drain and from the drain to the source.

The physical argument discussed above also allows us to answer in simple terms a central question 
in this work: Is the heat equally dissipated in both electrodes? In general, the answer is no, as 
we proceed to explain. Let us assume that the transmission function is energy-dependent and, for 
instance, that the transmission is higher for energies in the lower part of the transport window, 
as shown in figure~\ref{fig_HD}(c). As we shall discuss below, this corresponds to a situation
where the transport is dominated by the highest occupied molecular orbital (HOMO). In this case, 
for a positive bias, it is obvious that a larger portion of the energy is dissipated in the source
because it is more probable that electrons with a smaller excess energy tunnel through the molecule
and release their energy in the drain. Obviously, if we reversed the transmission landscape, see
figure~\ref{fig_HD}(d), the heat would be preferentially dissipated downstream following the 
electron flow, \emph{i.e.}\ in the drain. To conclude this subsection, we would like to remark that 
these arguments also make clear the fact that, in general, heat dissipation in a given electrode
depends on the bias polarity, \emph{i.e.}\ it depends on the current direction.
 
\subsection{General considerations} \label{sec-GC}

As explained above, within the Landauer approach the heat dissipated in the electrodes of a 
nanoscale junction is fully determined by its transmission characteristics. From equations 
(\ref{eq-Qs}) and (\ref{eq-Qd}) one can draw some general conclusions about the symmetry
of the heat dissipation between electrodes and with respect to the bias polarity. Let us
first discuss under which circumstances heat is equally dissipated in both electrodes.
For this purpose, we assume from now on (without loss of generality) that all the energies in 
the problem are measured with respect to the equilibrium electrochemical potential of the system, which
we set to zero ($\mu=0$). Moreover, we assume that the electrochemical potentials are shifted 
symmetrically with the bias voltage, \emph{i.e.}\ $\mu_S = eV/2$ and $\mu_D = -eV/2$. This means 
in practice that the energy $E$ is measured with respect to the center of the transport window. 
With this choice, the power dissipated in the source electrode is given by
\begin{equation}
\label{eq-Qssym}
Q_S(V) = \frac{2}{h} \int^{\infty}_{-\infty} (eV/2 -E) \tau(E,V) [ f(E-eV/2) - f(E+eV/2) ] dE, \\
\end{equation}
where $f(E) = 1/[1 + \exp(E/k_{\rm B}T)]$ is the Fermi function. Using the change of variable 
$E \to -E$, the previous expression becomes
\begin{equation}
Q_S(V) = \frac{2}{h} \int^{\infty}_{-\infty} (E+eV/2) \tau(-E,V) [ f(E-eV/2) - f(E+eV/2) ] dE , \\
\end{equation}
where we have used the relation $f(-E) = 1 - f(E)$. This latter equation must be compared with the 
corresponding expression for the power dissipated in the drain, which now reads
\begin{eqnarray}
\label{eq-Qdsym}
Q_D(V) = \frac{2}{h} \int^{\infty}_{-\infty} (E + eV/2) \tau(E,V) [ f(E-eV/2) - f(E+eV/2) ] dE .
\end{eqnarray}
Thus, we conclude that a sufficient condition to have equal heat dissipation in both electrodes,
\emph{i.e.}\ $Q_S(V) = Q_D(V)$, is that the transmission function fulfills that $\tau(E,V) = \tau(-E,V)$.
This means that if the transport is electron-hole symmetric, then heat is equally dissipated in the
source and in the drain. A simple corollary of this condition is that if the transmission is 
energy-independent in the transport window, then the power dissipation is the same in both electrodes.
In particular, if the junction is ballistic, \emph{i.e.}\ if $\tau(E,V) = M$, $M$ being an integer 
equal to the number of open conduction channels in the system, then heat is equally dissipated 
in both electrodes. Notice that these conclusions can be also drawn from the handwaving arguments
explained at the end of the previous subsection.

Another crucial question that can be answered in general terms is: Does the power dissipated in a given
electrode depend on the bias polarity (or current direction)? With manipulations similar to those
of the previous paragraph, it is straighforward to show that $Q_{S,D}(V) = Q_{S,D}(-V)$ if $\tau(E,V)
= \tau(-E,-V)$ is satisfied. This means, in particular, that if the transmission does not depend 
significantly on the bias voltage, the power dissipated in an electrode is independent of the bias
polarity as long as there is electron-hole symmetry. Obviously, if the transmission is independent 
of both the energy and the bias, then the heating is symmetric with respect to the inversion of
the bias polarity. Again, this is what occurs in any ballistic structure.

On the other hand, one can also show that the relation $Q_{S,D}(V) = Q_{D,S}(-V)$ is satisfied if
$\tau(E,V) = \tau(E,-V)$. This relation implies the following one: $Q_{S,D}(V) + Q_{S,D}(-V)
= Q_S(V) + Q_D(V) = Q_{\rm Total}(V) = IV$. This means that if the transmission is symmetric with 
respect to the inversion of the bias voltage, then the sum of the dissipated powers for 
positive and negative bias in a given electrode is equal to the total power dissipated in the junction. 
This relation is then expected to hold for left-right symmetric junctions, where the voltage profile
is inversion symmetric with respect to the center of the junction.

From the general considerations above, it is clear that in order to have a heating asymmetry of the type
$Q_S(V) \neq Q_D(V)$ or $Q_{S,D}(V) \neq Q_{S,D}(-V)$, one needs a certain degree of electron-hole 
asymmetry in the transmission function. This can be seen explicitly by expanding equations (\ref{eq-Qssym})
and (\ref{eq-Qdsym}) to first order in the bias voltage, which leads to the following expression
\begin{equation}
\label{eq-Qsd1st}
\fl Q_{S,D}(V) = \pm \frac{2eV}{h} \int^{\infty}_{-\infty} E \tau(E,V=0) \frac{\partial f(E,T)}
{\partial E} dE \, + \, O(V^2) = \pm GTSV \, + \, O(V^2) .
\end{equation}
Here, $G$ is the linear electrical conductance of the junction, $T$ is the absolute temperature, and 
$S$ is the thermopower or Seebeck coefficient of the junction. Let us remind that in the Landauer
approach $G$ and $S$ can be expressed in terms of the zero-bias transmission as
\begin{eqnarray}
G & = & \frac{2e^2}{h} \int^{\infty}_{-\infty} \tau(E,V=0) [ - \partial f(E,T)/
\partial E] \; dE , \\
S & = & -\frac{1}{eT} \frac{ \int^{\infty}_{-\infty} E \tau(E,V=0) [-\partial f(E,T)/
\partial E] \; dE} {\int^{\infty}_{-\infty} \tau(E,V=0) [-\partial f(E,T)/ \partial E] \; dE}.
\end{eqnarray}

Equation (\ref{eq-Qsd1st}) can also be written in the familiar form: $Q_{S,D}(V) = \pm \Pi I$ 
\cite{Ashcroft1976}, where $\Pi = TS$ is the Peltier coefficient of the junction and $I=GV$ is the 
electrical current in the linear regime. Moreover, from equation (\ref{eq-Qsd1st}) we can conclude that 
\begin{eqnarray}
\label{eq-Qsdasym}
Q_{S}(V) - Q_{D}(V) & = & \pm 2GTSV + O(V^2) , \\
Q_{S,D}(V) - Q_{S,D}(-V) & = & \pm 2GTSV + O(V^3) ,
\label{eq-Qsdasym2}
\end{eqnarray}
\emph{i.e.}\ \emph{the heating asymmetries between electrodes and with respect to the bias polarity are
determined to first order in the bias by the Seebeck coefficient of the junction}.

On the other hand, using the low temperature expansion of $G$ and $S$
\begin{equation}
\label{eq-GSlinear}
G = \frac{2e^2}{h} \tau(E_{\rm F},V=0) \;\;\; \mbox{and} \;\;\; S = - \frac{\pi^2 k^2_{\rm B}T}{3e} 
\frac{\tau^{\prime}(E_{\rm F}, V=0)}{\tau(E_{\rm F}, V=0)} ,
\end{equation}
where $\tau^{\prime}(E_{\rm F}, V=0)$ is the energy derivative of the zero-bias transmission
at the Fermi energy, $E_{\rm F} = 0$, we arrive at the following expression for the linear voltage 
term of the power dissipations at low temperature
\begin{equation}
\label{eq-QsdlowTV}
Q_{S,D}(V) = \mp \left( \frac{2e}{h} \right) \frac{\pi^2}{3} (k_{\rm B}T)^2 
\tau^{\prime}(E_{\rm F}, V=0) V \, + \, O(V^2).
\end{equation}
Thus, we see that the slope of the transmission function at the Fermi energy determines not only
the thermopower, but also the magnitude and the sign of the heating asymmetry to the lowest order
in the bias.

Equation (\ref{eq-Qsd1st}) implies that at sufficiently low bias, and if the thermopower is finite,
one of the electrodes may be cooled down by the passage of an electrical current (Peltier effect). How 
low must the voltage be to observe this cooling effect? To answer this question we need to consider 
the quadratic term of $Q_{S,D}(V)$ in the bias. (Notice that a quadratic term must exist in order to 
satisfy energy conservation, $Q_S(V) + Q_D(V) = Q_{\rm Total}(V)$, since at low bias $Q_{\rm Total}(V) 
= GV^2$.) If for simplicity we ignore the bias depedence of the transmission function, it is easy to 
see that the second-order term of $Q_{S,D}(V)$ is given by $(1/2)GV^2$, which is equal to half of the 
total power in the linear regime. This second-order term dominates over the linear one, see equation 
(\ref{eq-Qsd1st}), when $|V| > 2T|S|$ (above this bias, each electrode is heated up by the current). 
Thus for instance, if we assume room temperature ($T = 300$ K) and a typical value of $|S| = 10$ 
$\mu$V/K \cite{Reddy2007}, then the second-order term dominates the contribution to the heating for 
voltages $|V| > 6$ mV. This means in practice that in most molecular junctions we expect Joule heating 
to dominate over Peltier cooling over a wide range of bias. Of course, at a sufficiently high 
bias, higher order terms in the voltage expansion (beyond the quadratic one) may give a significant 
contribution to the power dissipation in a given electrode. In any case, all those non-linear contributions 
can be taken into account by using the full expressions for the power dissipated, see equations 
(\ref{eq-Qssym}) and (\ref{eq-Qdsym}). 

\subsection{Some lessons from a single-level model} \label{sec-SLM}

In order to illustrate some of the ideas discussed in the previous subsections and to gain some further 
insight, we analyze in this subsection the heat dissipation in molecular junctions with the help 
of a single-level model, sometimes refer to as \emph{resonant tunneling model} \cite{Zotti2010,Cuevas2010}. 
In this model one assumes that transport through a molecular junction is dominated by a single molecular 
orbital (typically the HOMO or the LUMO) and the transmission function is given by
\begin{equation}
\label{eq-tauSLM}
\tau(E) = \frac{\Gamma^2}{(E-\epsilon_0)^2 + \Gamma^2}.
\end{equation}
Here, $\epsilon_0$ is the energetic position of the molecular level measured with respect to the Fermi 
energy (set to zero) and $\Gamma$ is the strength of the metal-molecule coupling (or equivalently the level 
broadening). For simplicity, we assume here that the junction is symmetric, \emph{i.e.}\ the molecule is 
equally coupled to both electrodes. Let us remind that in this model the parameter $\Gamma$ is taken as a 
constant (energy-independent), which means that local density of states in the metal electrodes is assumed 
to be constant in the energy range involved in the transport window. Let us also recall that in the symmetric 
case considered here, the level position does not depend on the bias and the transmission function is thus 
independent of the applied voltage.

Within this model, the power dissipated in the source and in the drain electrodes is obtained by substituting equation 
(\ref{eq-tauSLM}) into equations (\ref{eq-Qssym}) and (\ref{eq-Qdsym}). Since the transmission does not 
depend explicitly on the bias voltage, the following two relations are satisfied (see subsection \ref{sec-GC})
\begin{equation}
Q_S(V) = Q_D(-V) \;\;\; \mbox{and} \;\;\; Q_{S,D}(V) + Q_{S,D}(-V) = 
Q_{\rm Total}(V) = IV .
\end{equation}
The first relation tells us that the power dissipated in one of the electrodes can be obtained from the
power dissipated in the other one by simply inverting the bias. The second relation implies that the sum 
of the power dissipated for positive and negative bias in a given electrode is equal to the total power 
dissipated in the junction.

\begin{figure}[t]
\begin{center} \includegraphics[width=0.8\columnwidth,clip]{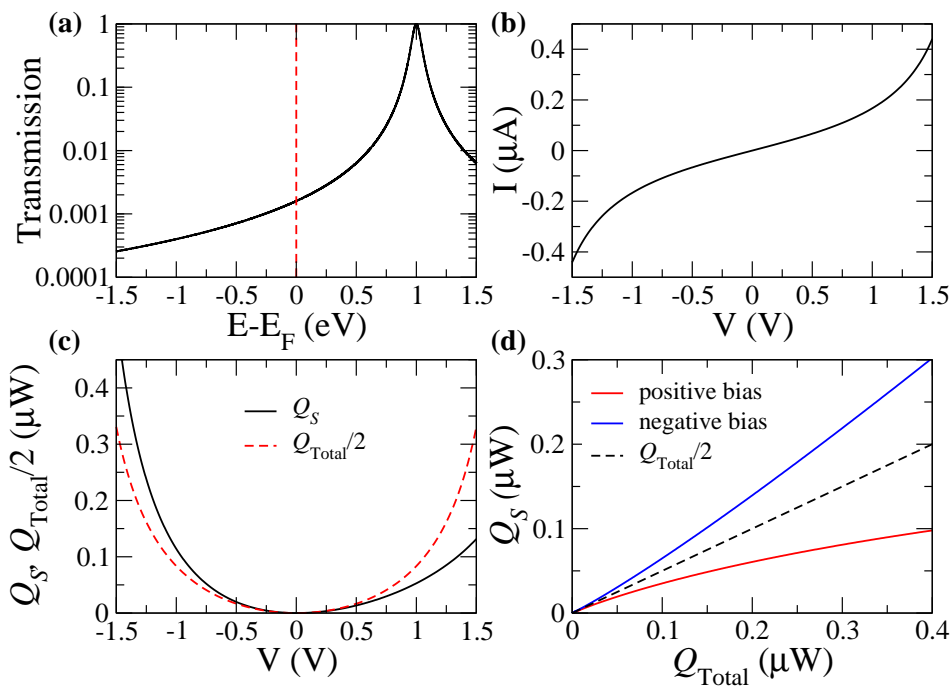} \end{center}
\caption{(a) The transmission as a function of energy for $\epsilon_0 = +1$ eV and $\Gamma = 40$ meV.
(b) The corresponding current-voltage characteristics. (c) Half of the total power dissipated in the
junction and the power dissipated in the source as a function of the bias. (d) The power dissipated in
the source as a function of the total power for positive and negative bias. The dashed line correspond
to the power dissipated in a symmetric situation, \emph{i.e.} $Q_S(V) = Q_{\rm Total}/2$. In panels
(b-d) the results were obtained assuming room temperature.}
\label{fig_SLM_LUMO}
\end{figure}

Let us now use this model to simulate a junction where the transport is dominated by the LUMO. For this 
purpose, we choose $\epsilon_0 = +1$ eV (level above the Fermi energy) and $\Gamma = 40$ meV. With these values 
the transmission at the Fermi energy is $\tau(E_{\rm F}) = 1.59 \times 10^{-3}$. In figure \ref{fig_SLM_LUMO} 
we show for this example the transmission as a function of energy (panel a), the current-voltage (I-V) characteristics
(panel b), the total power and power dissipated in the source as a function of the bias (panel c), and the 
power dissipated in the source as a function of the total power for both positive and negative bias (panel d). 
The voltage range in panels b and c has been chosen as to describe the typical regime explored in most experiments
in which the highest bias is not sufficient to reach the resonant condition $|eV| = 2 |\epsilon_0|$. 
The results of panels b-d have been calculated a room temperature (300 K), but in this off-resonant situation 
the corresponding results at zero temperature are similar (not shown here). The main conclusion of these 
results is that, as it can be clearly seen in panel d, the power dissipated in the source for negative bias is 
higher than for positive bias (the situation is the opposite in the drain). According to equation (\ref{eq-Qsdasym2}),
this fact is due to the negative value of the Seebeck coefficient, or equivalently to the positive slope of the
transmission at the Fermi energy, see equation (\ref{eq-QsdlowTV}). In this case, the non-linearities of
the I-V curve are reflected in the fact that the total power contains a significant quartic term ($\propto V^4$) 
and the bias asymmetry in the power dissipated in the source, $Q_S(V)-Q_S(-V)$, contains a significant cubic 
term ($\propto V^3$) and even higher order ones.

\begin{figure}[t]
\begin{center} \includegraphics[width=0.8\columnwidth,clip]{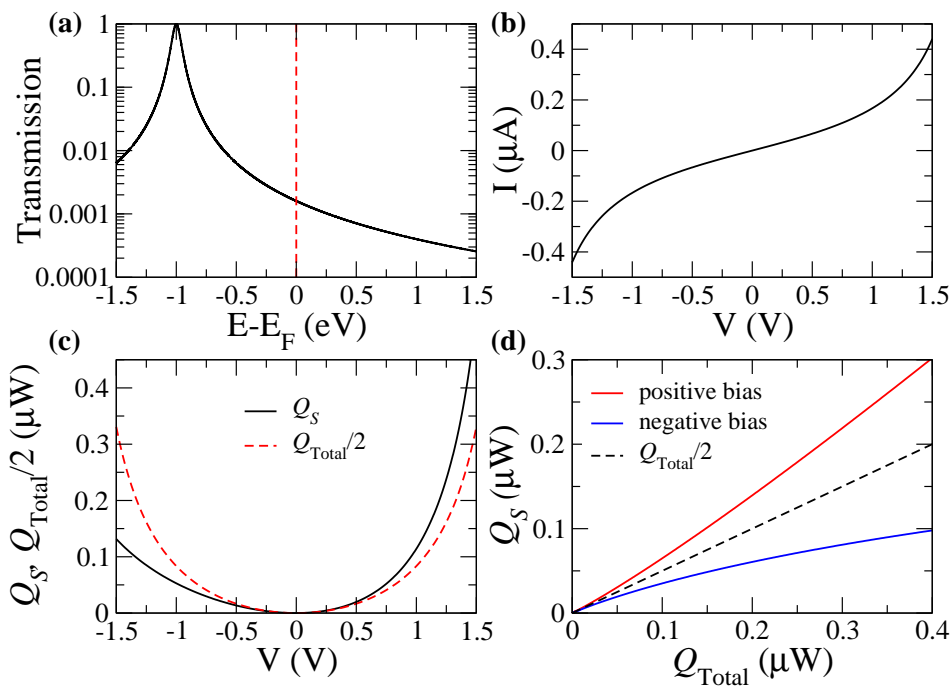} \end{center}
\caption{The same as in figure \ref{fig_SLM_LUMO} for $\epsilon_0 = -1$ eV.}
\label{fig_SLM_HOMO}
\end{figure}

To illustrate how the sign of the Seebeck coefficient is closely related to the asymmetries in the heat dissipation,
we now investigate a case where we just change the sign of $\epsilon_0$ with respect to the previous
example. This case corresponds to a situation where the transport is dominated by the HOMO. The corresponding
results are shown in figure \ref{fig_SLM_HOMO}. As one can see, while the I-V curve is exactly the same as
in the previous case, now a higher power is dissipated in the source for positive bias contrary to the 
example in figure \ref{fig_SLM_LUMO}. These results nicely illustrate how the heat dissipation can be
controlled by tuning the thermopower of the junctions: The power dissipation is always higher in the electrode 
with the highest electrochemical potential if the level lies below the Fermi level (in equilibrium), while it 
is higher in the electrode with the lowest electrochemical potential if the level is above it.

\begin{figure}[t]
\begin{center} \includegraphics[width=\columnwidth,clip]{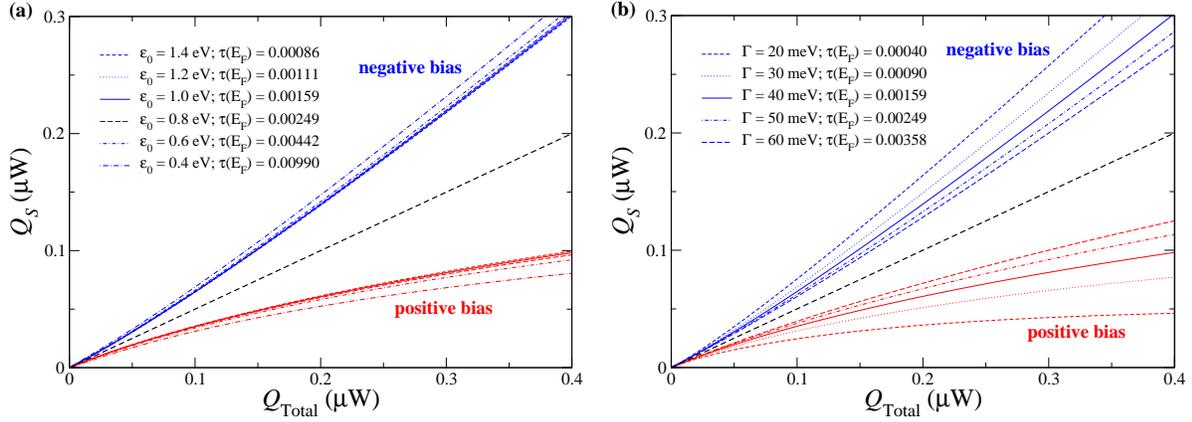} \end{center}
\caption{(a) The $Q_S$-$Q_{\rm Total}$ relation is shown for $\Gamma = 40$ meV and different values of the 
level position, as indicated in the legend. (b) The same relation for $\epsilon_0 = 1$ eV and different 
values of the coupling strength. In both panels, the corresponding values of the transmission at the Fermi 
energy are indicated in the legend. All the results were obtained at room temperature.}
\label{fig_SLM_eps}
\end{figure}

The relation between $Q_S$ and $Q_{\rm Total}$ is particularly useful to visualize the heating asymmetries
(both between electrodes and with respect to the bias polarity). In this sense, it is interesting to 
analyze to what extent this relation depends on variations of the parameters of the model. In this 
analysis we focus on the case of off-resonant transport, which is the most common situation in the
experiments. In figure \ref{fig_SLM_eps}(a) we show this relation for $\Gamma = 40$ meV and different 
values of the level position ranging from 0.4 to 1.4 eV. Notice that in all cases we are in the off-resonant 
regime ($|\epsilon_0| \gg \Gamma$). As one can see, the $Q_S$-$Q_{\rm Total}$ relation is rather insensitive 
to variations in the level position, in spite of the fact that the linear conductance changes by up to an 
order of magnitude.  In figure \ref{fig_SLM_eps}(b) we explore how this relation depends on the coupling 
strength. For this purpose, we have fixed the level position to $\epsilon_0 = +1$ eV and varied $\Gamma$ 
between 20 and 60 meV. Notice that the relation between the powers is more sensitive to variations in $\Gamma$, 
but again the variations are relatively small taking into account that the conductance has been varied by a 
factor of 10.

Now, let us try to shed some more light on the insensitivity of the $Q_S$-$Q_{\rm Total}$ relation to
variations of the junction parameters in an off-resonant transport situation. For this purpose, we have carried out
analytical calculations in the limit of zero temperature, which are also applicable to finite temperatures 
in an off-resonant situation. From equations (\ref{eq-Qssym}), (\ref{eq-Qdsym}), and (\ref{eq-tauSLM}), it 
is straightforward to show that
\begin{eqnarray}
\label{eq-Qs-SLM-an}
\fl \hspace{5mm} Q_S(V) & = & \frac{2}{h} \Gamma (eV/2-\epsilon_0) 
\left\{ \arctan \left[ \frac{eV/2-\epsilon_0}{\Gamma} \right]
+ \arctan \left[ \frac{eV/2+\epsilon_0}{\Gamma} \right] \right\} - \\ 
\fl & & \frac{\Gamma^2}{h} \left\{ \ln \left[ 1 + \left( \frac{eV/2-\epsilon_0}{\Gamma} \right)^2 \right]
- \ln \left[ 1 + \left( \frac{eV/2+\epsilon_0}{\Gamma} \right)^2 \right] \right\} , \nonumber \\
\label{eq-Qtotal-SLM-an}
\fl Q_{\rm Total}(V) & = & \frac{2e \Gamma V}{h} \left\{ \arctan \left[ \frac{eV/2-\epsilon_0}{\Gamma} \right]
+ \arctan \left[ \frac{eV/2+\epsilon_0}{\Gamma} \right] \right\} .
\end{eqnarray}
To establish the relation between these two powers, we first do a Taylor expansion in the bias of both 
expressions and focus on the off-resonant situation ($|\epsilon_0| \gg \Gamma$). Thus, the total power can 
be written in this limit as
\begin{equation}
Q_{\rm Total}(V) \approx GV^2 + \frac{e^2 G}{4\epsilon^2_0} V^4 + O(V^6) ,
\end{equation}
while the heating asymmetry adopts the form
\begin{equation}
Q_S(V) - Q_D(V) = Q_S(V) - Q_S(-V) \approx \frac{eG}{3\epsilon_0} V^3 + \frac{e^3G}{10 \epsilon^3_0} V^5 + O(V^7) .
\end{equation}
Here, $G$ is the linear conductance, which in the off-resonant limit reads $G \approx G_0(\Gamma / \epsilon_0)^2$, 
where $G_0=2e^2/h$ is the conductance quantum. If we now restrict ourselves to the low-bias limit where the total 
power is quadratic ($Q_{\rm Total}(V) \approx GV^2$), then the power dissipated in the source can be expressed as
$[Q_S = Q_{\rm Total}/2 + (Q_S-Q_D)/2]$
\begin{equation}
Q_S(Q_{\rm Total}) \approx \left\{ \begin{array}{cc}
\frac{1}{2} Q_{\rm Total} + \mbox{sgn}(\epsilon_0) \frac{e}{6G^{1/2}_0} \frac{1}{\Gamma} Q^{3/2}_{\rm Total} &
\mbox{(for negative bias)} \\
\frac{1}{2} Q_{\rm Total} - \mbox{sgn}(\epsilon_0) \frac{e}{6G^{1/2}_0} \frac{1}{\Gamma} Q^{3/2}_{\rm Total} &
\mbox{(for positive bias)} \end{array} \right. .
\end{equation}
This expression shows that the relation between $Q_S$ and $Q_{\rm Total}$ is independent of the level position in 
an off-resonant situation. Notice that the level position only enters via its sign (positive for LUMO-dominated 
cases and negative for HOMO-dominated ones). This analytical expression nicely illustrates the insensitivity
of this relation to the level position found in the numerical results above.

\section{Ab initio calculations of the transmission characteristics of molecular junctions} \label{sec-abinitio}

As we have seen in the previous section, within the Landauer approach all the transport properties
are determined by the transmission characteristics of the junctions. Thus, in order to describe 
realistic systems, we need microscopic methods to compute these characteristics. In this section 
we discuss the two \emph{ab initio} methods that we have employed in this work to compute the
transmission through molecular junctions and, in turn, to study the corresponding heat dissipation.

\subsection{Density functional theory}

The goal of this subsection is to describe the first \emph{ab initio} method, which is based on density functional 
theory (DFT). This DFT-based transport method has been described in great detail in Ref.~\cite{Pauly2008a} 
and therefore, our discussion here will be rather brief. 
This method is built upon the quantum-chemistry code TURBOMOLE \cite{Ahlrichs1989}. A central issue in our approach 
is the description of the electronic structure of the junctions within DFT. In all the calculations presented 
here we have used the BP86 exchange-correlation functional \cite{Becke1988,Perdew1986} and the Gaussian basis set def-SVP 
\cite{Schaefer1992}. The total energies were converged to a precision of better than $10^{-6}$ atomic units, and 
structure optimizations were carried out until the maximum norm of the Cartesian gradient fell below $10^{-4}$ atomic 
units. In what follows, we shall discuss the two main steps in our method: (i) construction of the geometries of 
the single-molecule junctions and (ii) determination of the electronic structure of the junctions and calculation 
of the transmission function using Green's function techniques.

In the construction of the junction geometries we proceed as follows. We first place the relaxed molecule in between 
two gold clusters with 20 (or 19) atoms. Subsequently, we perform a new geometry optimization by relaxing the 
positions of all the atoms in the molecule as well as the four (or three) gold atoms on each side that are 
closest to it, while the other gold atoms are kept fixed. Afterwards, the size of the gold cluster is increased 
to about 63 atoms on each side in order to describe correctly both the metal-molecule charge transfer and the energy 
level alignment. Finally, the central region, consisting of the molecule and one or two Au layers on each side, is 
coupled to ideal gold surfaces, which serve as infinite electrodes and are treated consistently with the same 
functional and basis set within DFT.

\begin{figure}[t]
\begin{center} \includegraphics[width=0.9\columnwidth,clip]{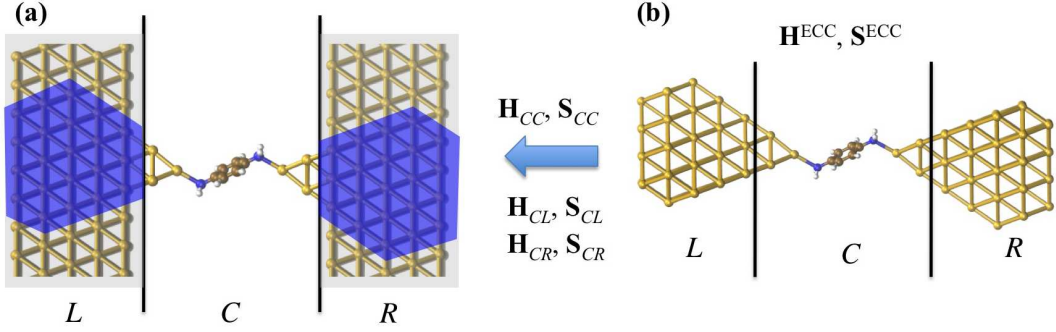} \end{center}
\caption{Schematic representation of the DFT-based transport method. The single-molecule contact (a) is divided 
into a $C$ region and two semi-infinite $L$ and $R$ electrodes. Using a similar division for the ECC (b), 
information on the electronic structure of the $C$ region (${\bf H}_{CC}, {\bf S}_{CC}$) as well as the 
$CL$ and $CR$ couplings (${\bf H}_{CL}, {\bf S}_{CL}$ and ${\bf H}_{CR}, {\bf S}_{CR}$) is extracted. 
The electrode surface Green's functions ${\bf g}^r_{LL} $ and ${\bf g}^r_{RR}$, needed for the
computation of the self-energies ${\bf \Sigma}^r_L$ and ${\bf \Sigma}^r_R $ [equation~(\ref{eq:Sigma_X})], 
are determined in a separate calculation \cite{Pauly2008a}.}
\label{fig_method_geom}
\end{figure}

To determine the electronic structure of the infinite junctions and to compute the transmission characteristics 
we make use of a combination of Green's function techniques and the Landauer formula expressed in a local nonorthogonal 
basis. Briefly, the local basis allows us to partition the basis states into $L$, $C$, and $R$ ones, according to 
the division of the contact geometry, see figure~\ref{fig_method_geom}. Thus, the Hamiltonian (or single-particle Fock)
matrix ${\bf H}$, as well as the overlap matrix ${\bf S}$, can be written in the block form
\begin{equation}
{\bf H}=\left(\begin{array}{ccc}
{\bf H}_{LL} & {\bf H}_{LC} & {\bf 0}\\ {\bf H}_{CL} & {\bf H}_{CC} & {\bf H}_{CR}\\
{\bf 0} & {\bf H}_{RC} & {\bf H}_{RR}\end{array}\right).
\label{eq:H}
\end{equation}
Within the Landauer approach, the low-temperature conductance is given by $G = G_0 \tau(E_{\rm F})$. It is
important to remark that we shall restrict ourselves to the case of zero bias throughout this discussion. 
The energy-dependent transmission $\tau(E)$ can be expressed in terms of the Green's functions as
\cite{Cuevas2010}
\begin{equation}
\tau(E) = \mathrm{Tr} \left[{\bf \Gamma}_{L} {\bf G}_{CC}^{r} {\bf \Gamma}_{R} {\bf G}_{CC}^{a} \right],
\label{eq:TE}
\end{equation}
where the retarded Green's function is given by
\begin{equation}
{\bf G}_{CC}^{r}(E) = \left[ E {\bf S}_{CC} - {\bf H}_{CC} - {\bf \Sigma}_{L}^{r}(E) - 
{\bf \Sigma}_{R}^{r}(E)\right]^{-1} ,
\label{eq:G_CC}
\end{equation}
and ${\bf G}_{CC}^{a} = \left[{\bf G}_{CC}^{r}\right]^{\dagger}$. The self-energies in the previous
equation adopt the form
\begin{equation}
{\bf \Sigma}_{X}^{r}(E) = \left({\bf H}_{CX} - E {\bf S}_{CX} \right) {\bf g}_{XX}^{r}(E) \left({\bf H}_{XC} 
- E {\bf S}_{XC} \right).
\label{eq:Sigma_X}
\end{equation}
On the other hand, the scattering rate matrices that enter the expression of the transmission are given 
by ${\bf \Gamma}_{X}(E) = -2\mathrm{Im} \left[{\bf \Sigma}_{X}^{r}(E) \right]$, and ${\bf g}_{XX}^{r}(E)=
(E {\bf S}_{XX} - {\bf H}_{XX})^{-1}$ are the electrode Green's functions with $X=L, R$.

In order to describe the transport through the contact shown in figure~\ref{fig_method_geom}(a), we first 
extract ${\bf H}_{CC}$ and ${\bf S}_{CC}$ and the matrices ${\bf H}_{CX}$ and ${\bf S}_{CX}$ from a DFT 
calculation of the extended center cluster (ECC) in figure~\ref{fig_method_geom}(b). The blue-shaded atoms 
in regions $L$ and $R$ of figure~\ref{fig_method_geom}(a) are assumed to be those coupled to the $C$ region. 
Thus, ${\bf H}_{CX}$ and ${\bf S}_{CX}$, obtained from the ECC, serve as the couplings to the electrodes in 
the construction of ${\bf \Sigma}_{X}^{r}(E)$.

On the other hand, the electrode Green's functions ${\bf g}_{XX}^{r}(E)$ in equation~(\ref{eq:Sigma_X}) are 
modeled as surface Green's functions of ideal semi-infinite crystals. To obtain these Green's functions,
we first compute separately the electronic structure of a spherical fcc gold cluster with 429 atoms. Second,
we extract the Hamiltonian and overlap matrix elements connecting the atom in the origin of the cluster 
with all its neighbors. Then, we use these bulk parameters to model a semi-infinite crystal 
which is infinitely extended perpendicular to the transport direction. The surface Green's functions are 
then calculated from this crystal with the help of the decimation technique \cite{Guinea1983,Pauly2008a}. 
In this way we describe the whole system consistently within DFT, using the same nonorthogonal basis set
and exchange-correlation functional everywhere. 

\subsection{DFT+$\Sigma$} \label{sec-DFTSigma}

In order to cross-check the validity of our conclusions on heat dissipation in single-molecule
junctions, we have employed a second method to compute the transmission characteristics. This method 
is known as self-energy corrected density functional theory DFT+$\Sigma$ and it is an extension of 
the DFT method that has been introduced to cure some of its known deficiencies \cite{Quek2007}. It 
is well-known that due to self-interaction errors in the standard exchange-correlation functionals and 
image charge effects, DFT-based methods have difficulties to accurately describe the energy gap and level 
alignment of molecules at surfaces \cite{Garcia-Lastra2009}. In particular, DFT tends to underestimate
the HOMO-LUMO gap, which in the context of molecular transport junctions implies that this
method tends to systematically overestimate the conductance. The DFT+$\Sigma$ approach, which we
describe in this subsection, has been shown to improve the agreement with the experiments for
both conductance and thermopower \cite{Mowbray2008,Quek2009,Quek2011,Widawsky2011}.

In our implementation of the DFT+$\Sigma$ method we have followed Ref.~\cite{Mowbray2008} (see also 
Ref.~\cite{Markussen2013}). This method aims at constructing accurate quasiparticle energies and 
starts by correcting the DFT-based energy levels of the gas phase molecules. To be precise, the 
HOMO energy is shifted such that it corresponds to the negative ionization potential (IP), while 
the LUMO is shifted to agree with the negative electron affinity (EA). All other energies of occupied 
(unoccupied) levels are shifted uniformly by the same amount as the HOMO (LUMO). The IP and EA are
computed within DFT from total energy calculations as follows
\begin{equation}
\label{eq-IP-EA}
\mbox{IP} = E(Q=+e) - E(Q=0) \;\;\; \mbox{and} \;\;\; \mbox{EA} = E(Q=0) - E(Q=-e) ,
\end{equation}
where $E(Q=0)$ is the total energy of the neutral molecule and $E(Q=+e)$ ($E(Q=-e)$) is the 
total energy of the molecule with one electron removed (or added). 

These corrected levels are, in turn, shifted when the molecule is brought into the junction.
In particular, image charge interactions shift the energy of the occupied states up in
energy and the virtual (unoccupied) states down in energy \cite{Neaton2006}. The key idea
in the DFT+$\Sigma$ method is that the screening of the metallic electrodes can be described
classically as the interaction of point charges (in the molecule) with two perfectly 
conducting infinite surfaces. The idea goes as follows. First, from the Hamiltonian
${\bf H}_{CC}$ of the ECC, see equation (\ref{eq:H}), and the corresponding overlap matrix 
${\bf S}_{CC}$, we extract the matrices ${\bf H}_{mol}$ and ${\bf S}_{mol}$ corresponding 
to the basis functions on the molecular atoms. Then, this molecular Hamiltonian is
diagonalized, ${\bf H}_{mol} \vec{c}_m = \epsilon_m {\bf S}_{mol} \vec{c}_m$, to obtain
the eigenenergies, $\epsilon_m$, and the eigenvectors, $\vec{c}_m$, for the molecule
in the junction. Now, this information is used to construct a point charge distribution
associated to a given molecular orbital $m$: $\rho_m(\vec{r}) = \sum^N_{i=1} Q^m_i 
\delta(\vec{r} - \vec{r}_i)$, where the $N$ point charges $Q^m_i$ are located at the
positions $\vec{r}_i$ of the atoms of the molecule in the junction. The contribution
$Q^m_i$ for a given molecular orbital $m$ can be obtained from the Mulliken populations 
for the selected molecular orbital at a given atom $i$ as $Q^m_i = \sum_{\mu \in i} 
({\bf D}^m {\bf S}_{mol})_{\mu \mu}$, where the sum runs over the local atomic basis 
states $| \mu \rangle$ belonging to atom $i$ and ${\bf D}^m_{\mu \nu} = c_{\mu m} 
c_{\nu m}$ is the density matrix of molecular orbital $m$.

The potential energy, $\Delta_m$, of a point charge distribution $\rho_m(\vec{r})$
placed between two perfectly conducting planes located at $z=\pm L/2$ is given by
(in atomic units)
\begin{eqnarray}
\label{eq-imagpot}
\fl \Delta_m = \frac{1}{2} \sum^N_{i,j=1} Q^m_i Q^m_j & & \\
\fl \times \sum^{\infty}_{n=1} \left(
\frac{-1}{\sqrt{ [z_i + z_j + (2n-1)L ]^2 + |\vec{r}^{\parallel}_i - \vec{r}^{\parallel}_j|^2}} +
\frac{-1}{\sqrt{[z_i + z_j - (2n-1)L ]^2 + |\vec{r}^{\parallel}_i - \vec{r}^{\parallel}_j|^2}}
\right. && \nonumber \\ \fl \left.
+ \frac{1}{\sqrt{[z_i + z_j + 2nL ]^2 + |\vec{r}^{\parallel}_i - \vec{r}^{\parallel}_j|^2}} +
\frac{1}{\sqrt{[z_i + z_j - 2nL ]^2 + |\vec{r}^{\parallel}_i - \vec{r}^{\parallel}_j|^2}} \right), 
&& \nonumber
\end{eqnarray}
where $\vec{r}^{\parallel}$ is the component of $\vec{r} = \vec{r}^{\parallel} + z \hat z$
parallel to the planes.

Following Ref.~\cite{Markussen2013}, we use the charge distribution of the HOMO to calculate
the image charge correction, $\Delta_{\rm occ}=-\Delta_{\rm HOMO}$, for all the occupied states, whereas
we use the LUMO charge distribution to determine the correction $\Delta_{\rm virt}=\Delta_{\rm LUMO}$ for the
virtual or unoccupied states. The position of the image plane is choosen to be 1.47 $\textrm{\AA}$ 
outside of the first unrelaxed gold layer of the left and right electrode. This, however, constitutes 
an approximation as we neglect the screening introduced by the apex Au atoms and the position of 
1.47 $\textrm{\AA}$ is valid only for a perfectly flat Au surface \cite{Lam1993,Widawsky2011}.

So finally, the energy shift applied to all the occupied states is given by $\Sigma_{\rm occ}
= -\mbox{IP} - \epsilon_H + \Delta_{\rm occ}$, and the corresponding shift for the
unoccupied ones by $\Sigma_{\rm virt} = -\mbox{EA} - \epsilon_L + \Delta_{\rm virt}$.
Here, $\epsilon_H$ and $\epsilon_L$ are the Kohn-Sham HOMO and LUMO energies from the
gas-phase calculation. With these shifts we obtain a modified molecular Hamiltonian,
$\tilde {\bf H}_{mol}$, which replaces ${\bf H}_{mol}$ in ${\bf H}_{CC}$.  Then, the method 
proceeds exactly as in the DFT case to compute the transmission of the junctions.

\section{Heat dissipation in benzene-based single-molecule junctions} \label{sec-MJ}

In this section we discuss the results obtained with the \emph{ab initio} methods described
in the previous section for the heat dissipation and its relation to thermopower in 
single-molecule junctions based on benzene derivatives. To be precise, we shall analyze
benzene molecules attached to gold electrodes via four common anchoring groups in molecular 
electronics: amine (-NH$_2$), isonitrile (-NC), nitrile (-CN), and thiol (-SH). These four 
groups have different character (electron-donating vs.\ electron-withdrawing) 
\cite{Hansch1991,Clayden2012} and they will help us to illustrate that the concepts used to 
tune the thermopower are very similar to those that govern the heating asymmetries.
Moreover, two of these molecules were investigated in our heat dissipation experiments of 
Ref.~\cite{Lee2013}, which allows us to gauge the quality of the theoretical results, and for 
some of them the conductance and the thermopower have been reported experimentally, as we 
discuss below. 

\begin{figure}[b]
\begin{center} \includegraphics[width=0.8\columnwidth,clip]{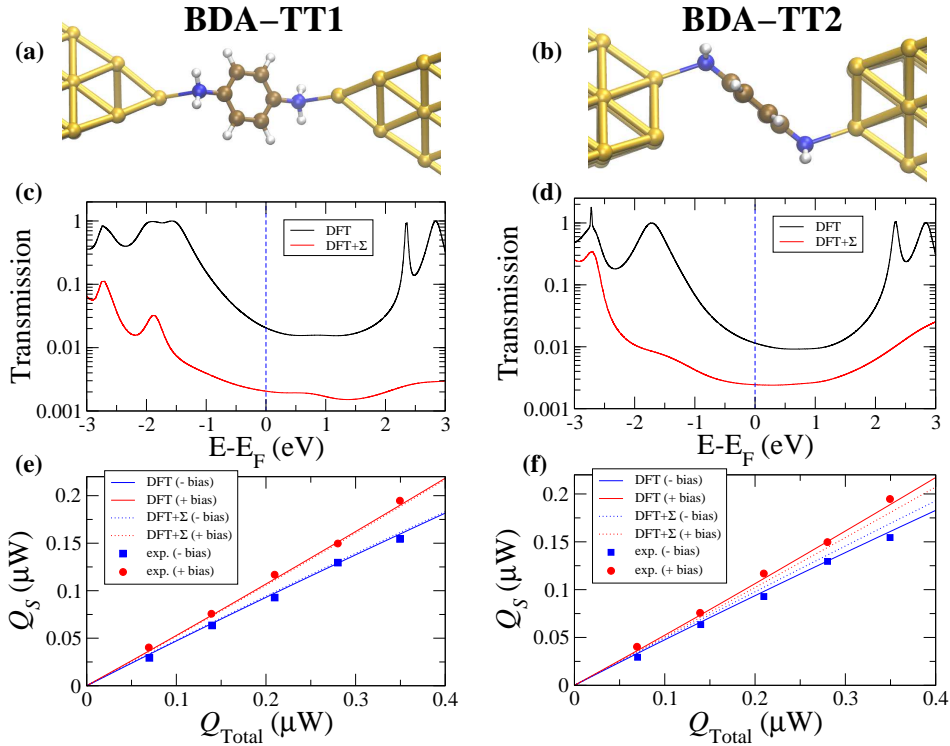} \end{center}
\caption{\emph{Ab initio} results for the transmission and heat dissipation in Au-BDA-Au junctions.
(a-b) The two atop Au-BDA-Au junction geometries investigated here. (c-d) The computed zero-bias
transmission as a function of the energy for the two geometries. We show the results obtained
with the two \emph{ab initio} methods discussed in the text: DFT and DFT+$\Sigma$. For clarity, the
position of the Fermi level is indicated with vertical lines. (e-f) The corresponding
results for the power dissipated in the source electrode, for positive and negative biases, as
a function of the total power dissipated in the two Au-BDA-Au junctions. The symbols correspond to
the experimental results of Ref.~\cite{Lee2013}.}
\label{fig_heat_BDA}
\end{figure}

Let us start our discussion with the case of benzenediamine (BDA). We have first explored different
binding geometries for this molecule between gold electrodes and found that the amine group only
binds to undercoordinated Au sites, in agreement with previous studies \cite{Venkataraman2006,Quek2007,Li2007}.
We show two examples of atop geometries for the Au-BDA-Au junctions in figures \ref{fig_heat_BDA}(a)
and (b), which will be referred to as TT1 and TT2. These geometries are potential candidates to
describe the configurations that give rise to the peaks in the experimental conductance histograms
\cite{Venkataraman2006,Lee2013}. The corresponding results for these two geometries for the zero-bias
transmission as a function of energy are shown in figures \ref{fig_heat_BDA}(c) and (d). We show the
results obtained with the two methods described in the previous section, namely DFT and DFT+$\Sigma$.
The values of the DFT conductances are equal to $2.1 \times 10^{-2}G_0$ (TT1) and $1.1 \times 10^{-2}G_0$
(TT2), which are slightly higher than the preferential experimental value of $6.4 \times 10^{-3}G_0$
reported in Ref.~\cite{Venkataraman2006} or $5.0 \times 10^{-3}G_0$ found in Ref.~\cite{Lee2013}.
As mentioned in the previous section, this typical overestimation of the conductance is attributed to
exchange and correlations effects that are not properly described within the DFT framework with conventional
exchange-correlation functionals. Such functionals tend to underestimate the HOMO-LUMO gap and
therefore they typically overestimate the conductance. On the other hand, the transmission curves indicate
that the HOMO of the molecule dominates the charge transport in these junctions, which results in a
negative slope at the Fermi energy and, in turn, in a positive Seebeck coefficient. Using equation
(\ref{eq-GSlinear}), we obtained a value for the thermopower at room temperature (300 K) of 7.2
$\mu$V/K for TT1 and 5.2 $\mu$V/K for TT2, which are similar to other DFT calculations \cite{Markussen2013}.
These values have to be compared with the experimental value of 2.3 $\mu$V/K reported in Ref.~\cite{Malen2009}.
In the case of the DFT+$\Sigma$ method, the transmission curves exhibit a much larger HOMO-LUMO gap, as compared
to DFT, and the values at the Fermi energy are around an order of magnitude smaller. As explained
in subsection \ref{sec-DFTSigma}, this is due to the significant shift introduced in this method in
all the occupied and empty states, see table~\ref{table1}. To be precise, the conductance values within
this method are equal to $2.1 \times 10^{-3}G_0$ for TT1 and $2.4 \times 10^{-3}G_0$ for TT2.
These results underestimate the experimental results by a factor 2-3. With respect to the thermopower,
the values are in this case 2.2 and 0.8 $\mu$V/K for TT1 and TT2, respectively. The values of the
conductance and the thermopower for the different molecular junctions discussed in the text are
summarized in table~\ref{table2}.

\begin{table}[t]
\caption{\label{table1}Relevant energies in the DFT+$\Sigma$ method for the different molecular 
junctions discussed in the text. Here, $-\mbox{IP}-\epsilon_H$ and $-\mbox{EA}-\epsilon_L$ correspond to 
the energy shifts of the occupied and unoccupied states, respectively, of the molecules in gas phase
as determined from the HOMO and LUMO. $\Delta_{\rm occ}$ and $\Delta_{\rm virt}$ correspond to the image 
charge energies for the occupied and unoccupied states, respectively. $\Sigma_{\rm occ}$ ($\Sigma_{\rm virt}$) 
is the total energy shift applied to the occupied (unoccupied) states of the molecule in the junction. 
All the energies are in units of eV.}
\begin{indented}
\item[]\begin{tabular}{@{}ccccccc}
\br
 & $-\mbox{IP}-\epsilon_H$ & $-\mbox{EA}-\epsilon_L$ & $\Delta_{\rm occ}$ &
$\Delta_{\rm virt}$ & $\Sigma_{\rm occ}$ & $\Sigma_{\rm virt}$ \\
\mr
{\bf BDA-TT1} & $-2.71$ & $2.70$ & 0.65 & $-0.65$ & $-2.06$ & 2.05 \\
\mr
{\bf BDA-TT2} & $-2.71$ & $2.70$ & 0.99 & $-0.94$ & $-1.72$ & 1.76 \\
\mr
{\bf BDNC-TT1} & $-2.53$ & $2.39$ & 0.62 & $-0.65$ & $-1.91$ & 1.74 \\
\mr
{\bf BDNC-TT2} & $-2.52$ & $2.37$ & 0.78 & $-0.84$ & $-1.74$ & 1.53 \\
\mr
{\bf BDT} & $-2.49$ & $2.58$ & 0.98 & $-0.93$ & $-1.51$ & 1.65 \\
\mr
{\bf BDCN} & $-2.53$ & $2.50$ & 0.85 & $-0.85$ & $-1.68$ & 1.65 \\
\br
\end{tabular}
\end{indented}
\end{table}
\begin{table}[b]
\caption{\label{table2}Conductance and thermopower (300 K) for the different molecular
junctions discussed in text and computed with both the DFT and DFT+$\Sigma$ method.}
\begin{indented}
\item[]\begin{tabular}{@{}ccccc}
\br
 & $G$ ($G_0$) & $S$ ($\mu$V/K) & $G$ ($G_0$) & $S$ ($\mu$V/K) \\
 & [DFT] & [DFT] & [DFT+$\Sigma$] & [DFT+$\Sigma$] \\
\mr
{\bf BDA-TT1} & $2.1 \times 10^{-2}$ & 7.2 & $2.1 \times 10^{-3}$ & 2.2 \\
\mr
{\bf BDA-TT2} & $1.1 \times 10^{-2}$ & 5.2 & $2.4 \times 10^{-3}$ & 0.8 \\
\mr
{\bf BDNC-TT1} & $4.8 \times 10^{-2}$ & $-29.0$ & $6.2 \times 10^{-3}$ & $-8.1$ \\
\mr
{\bf BDNC-TT2} & $1.4 \times 10^{-1}$ & $-48.2$ & $3.0 \times 10^{-3}$ & $-13.3$ \\
\mr
{\bf BDT} & $3.9 \times 10^{-2}$ & $6.4$ & $1.2 \times 10^{-2}$ & $2.4$ \\
\mr
{\bf BDCN} & $3.4 \times 10^{-1}$ & $-69.9$ & $1.9 \times 10^{-3}$ & $-11.5$ \\
\br
\end{tabular}
\end{indented}
\end{table}

On the other hand, making use of equations (\ref{eq-Qs}) and (\ref{eq-Qtotal}), we have computed 
the power dissipated in the source electrode for these two junctions as a function of the total power 
dissipated in the junction and the results for positive and negative bias are shown in figures 
\ref{fig_heat_BDA}(e) and (f). In these panels we have also included the experimental results of Ref.~\cite{Lee2013}.
It is worth stressing that in these calculations we have approximated the transmission curves by the 
zero-bias functions shown in figures \ref{fig_heat_BDA}(c) and (d). This approximation is justified 
by the fact that the highest bias explored in the experiments is still low as compared to the HOMO-LUMO 
gap of the molecules. Moreover, in our relatively weakly-coupled and symmetric molecular junctions the 
voltage is expected to drop mainly at the metal-molecule interfaces. This means that the relevant orbitals 
in the molecule are not significantly shifted by the bias and thus, the transmission is not expected 
to vary appreciably with voltage. This has been shown explicitly for similar molecules, for instance, in 
Refs.~\cite{Taylor2002,Darancet2012}. Let us remind that within our approximation, the power dissipated
in the drain electrode is given by $Q_D(V) = Q_S(-V)$. As one can see in figures \ref{fig_heat_BDA}(e) 
and (f), both methods, DFT and DFT+$\Sigma$, give very similar results for the relation between the power 
dissipations in spite of the different results that they produce for the conductance. Notice also that 
the results for both junction geometries are very similar and, more importantly, all the theoretical 
results are in good agreement with the experimental results of Ref.~\cite{Lee2013}. Furthermore, both 
the theoretical and the experimental results show that for this molecule there is more heat dissipation in 
the source electrode for positive bias, while the situation is reversed for negative bias, where a higher 
power is dissipated in the drain electrode. As explained in subsections \ref{sec-GC} and \ref{sec-SLM}, 
this behavior is due to the fact that the slope of the transmission at the Fermi energy is negative, 
which results in a positive Seebeck coefficient. On the other hand, the fact that the DFT and DFT+$\Sigma$ 
results agree and that the relation between the power disspations does not depend significantly on 
the junction geometry is a nice illustration of one of the central conclusions in subsection \ref{sec-SLM}, 
namely that this relation is quite insensitive to the level alignment in off-resonant situations 
like the one realized in these Au-BDA-Au junctions.

\begin{figure}[b]
\begin{center} \includegraphics[width=0.8\columnwidth,clip]{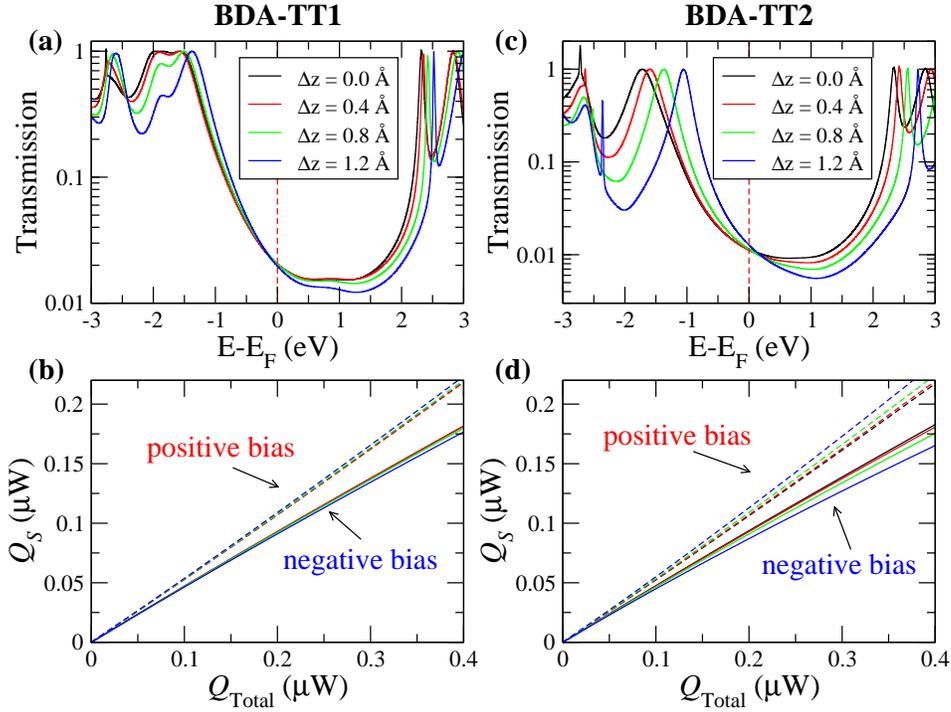} \end{center}
\caption{(a) Zero-bias transmission as a function of energy for Au-BDA-Au junctions obtained
by stretching the geometry of figure \ref{fig_heat_BDA}(a) by a distance $\Delta z$. (b) The
corresponding power dissipated in the source as a function of the total power. (c-d) The same 
as in panels (a-b) for geometries obtained by stretching the geometry of figure \ref{fig_heat_BDA}(b).
All the results shown in this figure were obtained with the DFT-based method.}
\label{fig_heat_BDA_stretching}
\end{figure}

To further illustrate the robustness of the conclusions drawn above, we have also explored how the heat 
dissipation evolves with the stretching of the contacts. For this purpose, we have started with the 
optimized geometries of \ref{fig_heat_BDA}(a) and (b) and we have separated step-wise the electrodes 
by a distance $\Delta z$ and in every step we have re-optimized the geometries. This process simulates
the junction stretching in typical break-junction experiments. In figure \ref{fig_heat_BDA_stretching} 
we show both the transmission curves and the corresponding heat dissipations, calculated with the 
DFT-based method, for a series of geometries obtained by elongating the original TT1 and TT2 geometries 
by up to 1.2 \AA. The important thing to remark here is that the relation between the power dissipations 
is hardly affected by the elongation process in the range of total powers explored in the experiments 
of Ref.~\cite{Lee2013}. Again, such an insensitivity can be attributed to the fact that we are in an
off-resonant situation, as explained in subsection \ref{sec-SLM}.

\begin{figure}[b]
\begin{center} \includegraphics[width=0.8\columnwidth,clip]{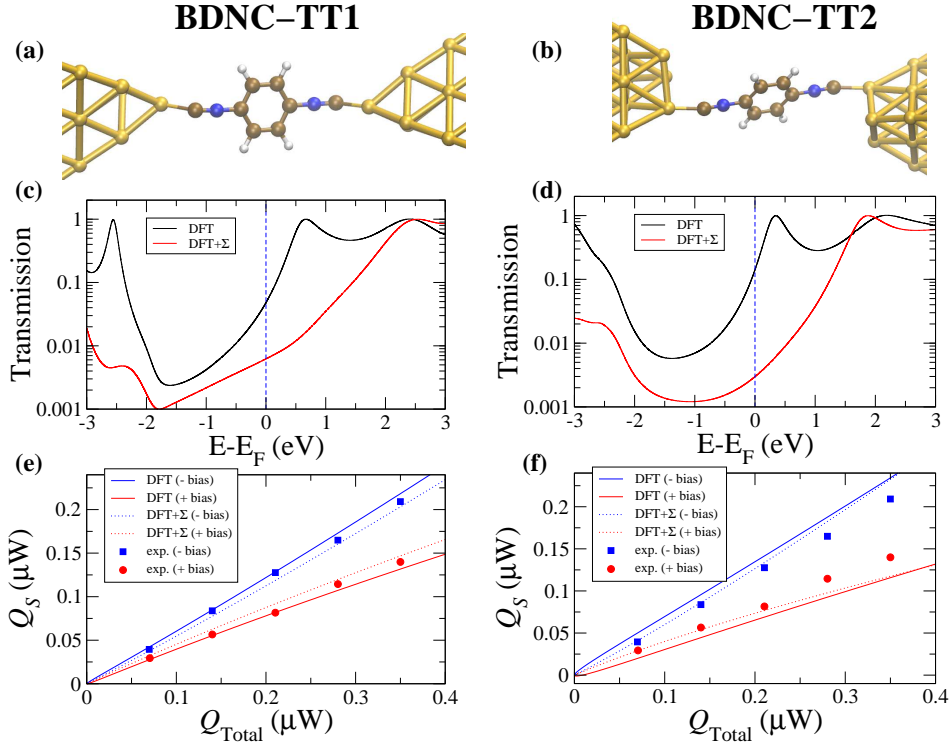} \end{center}
\caption{\emph{Ab initio} results for the transmission and heat dissipation in Au-BDNC-Au junctions.
(a-b) The two atop Au-BDNC-Au junction geometries investigated here. (c-d) The computed zero-bias
transmission as a function of the energy for the two geometries. We show the results obtained
with the two \emph{ab initio} methods discussed in the text: DFT and DFT+$\Sigma$. For clarity, the
position of the Fermi level is indicated by vertical dashed lines. (e-f) The corresponding
results for the power dissipated in the source electrode, for positive and negative biases, as a function
of the total power dissipated in the two junctions. The symbols correspond to the experimental
results of Ref.~\cite{Lee2013}.}
\label{fig_heat_BDNC}
\end{figure}

The heating asymmetries found in the Au-BDA-Au junctions can, in simple terms, be attributed to the
electron-donating \cite{Hansch1991,Clayden2012} character of the amine anchoring group, which results 
in HOMO dominated transport (or hole transport). In order to prove that the character of the terminal 
group determines both the thermopower and the heat dissipation, we now discuss the case of benzenediisonitrile 
(BDNC), since the isonitrile group has a well-known electron-withdrawing character \cite{Hansch1991,Clayden2012}.
Following the same analysis as for BDA, we have first investigated the most probable geometries of 
the Au-BDNC-Au junctions and we have found that the isonitrile group binds also preferentially to single
low-coordinated gold atoms in atop positions. This is in agreement with previous studies of adsorption 
of isocyanides on gold surfaces \cite{Gilman2008}. Two representative examples of the atop geometries 
found in our analysis are shown in figures \ref{fig_heat_BDNC}(a) and (b), from now on referred to as
TT1 and TT2. Notice that in both cases the C atom is directly bound to a single Au atom and the main 
difference between the two geometries lies in the shape of the electrodes. The results for the zero-bias 
transmission as a function of the energy for these two junctions are shown in figures \ref{fig_heat_BDNC}(c) 
and (d). For both geometries, and irrespective of the employed method, the low-bias transport is dominated 
by the LUMO. This is in qualitative agreement with previous results \cite{Xue2004,Dahlke2004}. The DFT
conductance values in these two examples are $4.8 \times 10^{-2}G_0$ (TT1) and 0.14$G_0$ (TT2). These values 
are clearly higher than the preferential value of $3\times 10^{-3}G_0$ \cite{Kiguchi2006} and $2 \times
10^{-3}G_0$ \cite{Lee2013} found experimentally. Again, we attribute this discrepancy to the intrinsic 
deficiencies of the existent DFT functionals that tend to underestimate the HOMO-LUMO gap. In this
case, DFT+$\Sigma$ produces much more satisfactory results for the low-bias conductance, 
$6.2 \times 10^{-3}G_0$ for TT1 and $3.0 \times 10^{-3}G_0$ for TT2. With respect to the thermopower,
as one can see in table~\ref{table2}, in all cases we find a negative value as a consequence of
the fact that the transport is dominated by the LUMO of the molecule (electron transport). Notice
also that the DFT+$\Sigma$ method produces smaller values for the thermopower due to the smaller
slope of the transmission at the Fermi energy, which is due to the fact that within this method
the LUMO lies around 2 eV higher than in the DFT method, where the LUMO is located relatively close to
the Fermi energy. We are not aware of measurements of the thermopower for BDNC.

With respect to the heat dissipation, we show the corresponding results for BDNC in figures 
\ref{fig_heat_BDNC}(e) and (f). Let us emphasize that these results were obtained using the zero-bias 
transmissions of figures \ref{fig_heat_BDNC}(c) and (d). For completeness, we have also included the
experimental results of Ref.~\cite{Lee2013}. The main conclusions from these results are the following.
First, a higher power is dissipated in the source electrode for negative bias, in strong constrast
to the BDA case. This is due to the fact that the slope of the transmission at the Fermi energy is
positive, which leads to a negative value for the thermopower. Second, both methods produce similar 
results for the relation between the power dissipations, in spite of the big differences in their
low-bias conductances. Again, this is a consequence of the insensitivity of this relation 
to the exact level alignment in an off-resonant situation. Third, the theoretical results 
reproduce qualitatively the experimental findings. Fourth, the heating assymetries are more pronounced
than in the BDA case. This is due to the larger Seebeck coefficient in the case of BDNC.

\begin{figure}[t]
\begin{center} \includegraphics[width=0.8\columnwidth,clip]{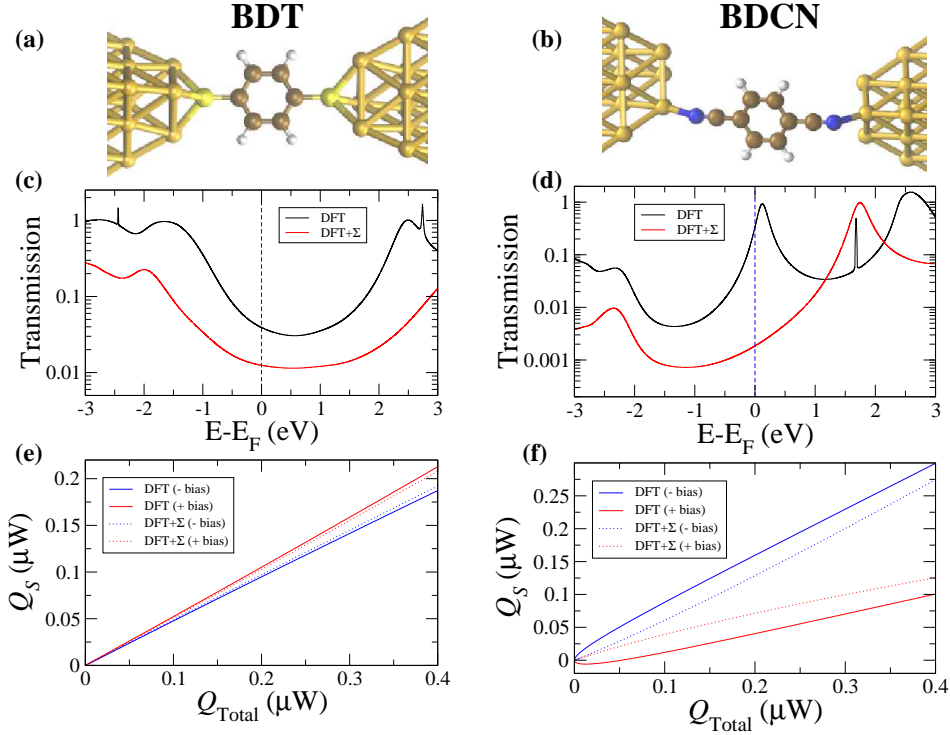} \end{center}
\caption{\emph{Ab initio} results for the transmission and heat dissipation in Au-BDT-Au and Au-BDCN-Au 
junctions. (a-b) Au-BDT-Au and Au-BDCN-Au junction geometries investigated here. (c-d) The computed 
zero-bias transmission as a function of the energy for the two junctions. We show the results obtained
with the two \emph{ab initio} methods discussed in the text: DFT and DFT+$\Sigma$. For clarity, the
position of the Fermi level is indicated by dashed vertical lines. (e-f) The corresponding results for 
the power dissipated in the source electrode, for positive and negative biases, as a function of the 
total power dissipated.}
\label{fig_heat_BDT-BDCN}
\end{figure}

The calculations discussed so far, and their agreement with the experiments, provide strong evidence 
of the relationship between the sign of the thermopower and the heating asymmetries. To provide
further evidence, we have also investigated single-molecule junctions based on benzenedithiol (BDT) 
and benzenedinitrile (BDCN). Following the classification in terms of Hammett constants 
\cite{Hansch1991,Clayden2012}, the amine group is strongly electron-donating, thiol is weakly 
electron-withdrawing, and nitrile and isonitrile are both strongly electron-withdrawing. In this 
sense, the naive expectation is that the thermoelectricity and the heat dissipation in Au-BDT-Au 
and Au-BDCN-Au junctions may be similar to those in Au-BDA-Au and Au-BDNC-Au, respectively. Let us 
show that this is indeed the case. In figure \ref{fig_heat_BDT-BDCN} we depict the results for the transmission 
functions and the power dissipations for the two geometries of the upper panels. The main features of 
these results are the following. In the case of BDT, transport is dominated by the HOMO, the thermopower 
is positive, and there is higher power dissipation in the source electrode for positive bias than for
negative bias. These results are indeed very similar to those obtained for the Au-BDA-Au junctions. 
The DFT transmission curve is similar to previous results in the literature \cite{Stokbro2003,
Mowbray2008,Pauly2008b}. As usual, the DFT+$\Sigma$ method \cite{note2,Garcia-Suarez2011} predicts a 
lower conductance value, which for this geometry is equal to $1.2 \times 10^{-2}G_0$, see table~\ref{table2}. 
This result is indeed very close to the experimental value of $1.1 \times 10^{-2}G_0$ reported in 
Ref.~\cite{Xiao2004}. With respect to the thermopower, both methods predict positive values that are 
close to the experimental one of 7.2 $\mu$V/K reported in Ref.~\cite{Baheti2008}.

The results for the Au-BDCN-Au junction of figure \ref{fig_heat_BDT-BDCN}(b) show that the transport in
this case is dominated by the LUMO, which leads to a negtive Seebeck coefficient and a higer power 
dissipation in the source electrode for negative bias, \emph{i.e.}\ very much like in the
case of Au-BDNC-Au junctions. Again, in this case the results obtained with DFT+$\Sigma$ predict
a much lower conductance than the DFT method, see table~\ref{table2}. In this case, the difference between
these methods for the power dissipation is more significant, as one can see in figure \ref{fig_heat_BDT-BDCN}(f).
The difference is particularly clear at low bias, where the DFT method predicts a small negative power
(cooling effect) for positive bias. This result is most likely an artifact of the position of the LUMO,
which lies very close to the Fermi energy resulting in a very large and negative thermopower,
see table~\ref{table2}. This cooling effect at relatively large total powers is absent in the 
DFT+$\Sigma$ result.

\section{Conclusions} \label{sec-conclusions}

We have presented a detailed theoretical analysis of the basic principles that govern heat
dissipation in single-molecule junctions. With the help of the Landauer approach, we have
shown how heat dissipation in the electrodes of a two-terminal molecular contact is 
determined by its transmission characteristics. In particular, we have shown how heat 
dissipation is, in general, different in the source and the drain electrodes and it
depends on the bias polarity or current direction. We have especially emphasized the
connection between these heating asymmetries and the thermopower of the junctions. As model
systems, we have studied single-molecule junctions based on benzene derivatives with different
anchoring groups. Making use of DFT and DFT+$\Sigma$ electronic structure methods, in combination 
with the Landauer approach, we have shown that both the heat dissipation and the thermopower 
can be chemically tuned by means of an appropriate choice of the terminal group. Moreover, we
have shown that our results are in good agreement with our experiments of Ref.~\cite{Lee2013}.   

Our analysis also provides simple guidelines for future studies of related phenomena such as
the Peltier effect in molecular junctions \cite{Galperin2009,Karlstroem2011,Dubi2011}. In all 
the junctions studied in this work, Joule heating dominates over Peltier cooling over a wide
range of voltages, but this can change with an appropriate choice of molecules and electrodes. 
For instance, resonant situations where the thermopower can take very large values can lead to 
cooling. It would also be interesting to explore the heat dissipation in situations where quantum 
interferences are expected to give rise to anomalously large values of the thermopower 
\cite{Finch2009,Bergfield2009,Karlstroem2011}. Another interesting possiblity is to use 
semiconducting electrodes where the gap may help to remove hot electrons and thus to cool down 
one the electrodes, as it has been demonstrated in hybrid superconducting systems in mesoscopic 
physics \cite{Giazotto2006}. 

\ack

L.A.Z.\ acknowledges financial support from the Spanish MICINN through Grant No.\ FIS2010-21883. 
This work was partly supported by a FY2012 (P12501) Postdoctoral Fellowship for Foreign Researchers 
from the Japan Society for Promotion of Science (JSPS) and by a JSPS KAKENHI, \emph{i.e.}, ``Grant-in-Aid 
for JSPS Fellows", Grant No.\ 24$\cdot$02501. F.P.\ gratefully acknowledges funding through the 
Carl Zeiss Foundation and the Baden-W\"urttemberg Foundation. P.R.\ acknowledges support from 
DOE-BES through a grant from the Scanning Probe Microscopy Division under award No.\ DE-SC0004871 
and support from the NSF under award No.\ CBET 0844902 and DOE-BES as part of an EFRC at the 
University of Michigan under award No.\ DE-SC0000957.


\begin{thebibliography}{99}
\bibitem{note1}
Let us stress that throughout this work, we define the source electrode as the lead from which
electrons flow for a positive bias voltage (and zero temperature), while the drain electrode is
meant to be the lead into which electrons are injected. Notice that with our convention, for a 
negative bias (and zero temperature) electrons flow from the drain to the source.
\bibitem{note2}
We note that as compared to the nitrogen-based anchoring groups NH$_2$, CN, and NC, which stay 
intact when bonded to gold, the thiols lose their terminal hydrogen atoms. The shifts of the 
occupied and unoccupied electronic levels determined for gas phase BDT with and without terminal 
hydrogens are very similar. The values given for BDT in table~\ref{table1} are for the hydrogen 
terminated case in the gas phase and the one without hydrogen for the polarization in the junction.
\end{thebibliography}
\end{document}